\documentclass[prd,twocolumn,tightenlines,showpacs,nofootinbib,amsfonts,amssymb,amsmath]{revtex4}
\usepackage{graphicx}
\begin{document}
\newcommand{\etal}{{\it et al.}}
\newcommand{\bx}{{\bf x}}
\newcommand{\bn}{{\bf n}}
\newcommand{\bk}{{\bf k}}
\newcommand{\dd}{{\rm d}}
\newcommand{\dslash}{D\!\!\!\!/}
\def\ga{\mathrel{\raise.3ex\hbox{$>$\kern-.75em\lower1ex\hbox{$\sim$}}}}
\def\la{\mathrel{\raise.3ex\hbox{$<$\kern-.75em\lower1ex\hbox{$\sim$}}}}
\def\beq{\begin{equation}}
\def\eeq{\end{equation}}

\leftline{UMN--TH--3016/11}
\leftline{SU--ITP--11/39}

\vskip-2cm

\title{Stability analysis of 5D gravitational solutions with $N$ bulk scalar fields
}

\author{Tony Gherghetta$^{1,2}$ and Marco Peloso$^3$
}

\affiliation{
$^1$School of Physics, University of Melbourne,
Victoria 3010, Australia\\
$^2$Stanford Institute of Theoretical Physics, Stanford University,
Stanford, CA 94305, USA\\
$^3$School of Physics and Astronomy,
University of Minnesota, Minneapolis, MN 55455, USA\\
}
\vspace*{2cm}
\begin{abstract}

We study the stability of 5D gravitational solutions containing an arbitrary
number of scalar fields. A closed set of equations is derived which governs the
background and perturbations of $N$ scalar fields and the metric, for arbitrary 
bulk and boundary scalar potentials. In particular the effect of the energy-momentum 
tensor of the scalar fields on the geometry is fully taken into account, together with 
all the perturbations of the system. The equations are explicitly written as an 
eigenvalue problem, which can be readily solved to determine the stability of the 
system and obtain the properties of the fluctuations, such as masses and couplings.  
As an example, we study a dynamical soft-wall model with two bulk scalar fields 
used to model the hadron spectrum of QCD and the Higgs sector of electroweak 
physics. It is shown that there are no tachyonic modes, and that there is a 
(radion) mode whose mass is suppressed by a large logarithm compared to
that of the other Kaluza-Klein modes.
\end{abstract}
 \date{September  2011}
 \maketitle

\section{Introduction}

Brane-world models have proved to be very useful for model building, providing a way to not only address the hierarchy problem in the Standard Model~\cite{ArkaniHamed:1998rs, Antoniadis:1998ig,Randall:1999ee}, but 
also to explain the hierarchy of fermion masses and mixings~\cite{ArkaniHamed:1999dc, Grossman:1999ra, Gherghetta:2000qt, Huber:2000ie}. A variety of constructions are usually employed with fields, other than gravity, in the bulk and/or on the branes. Furthermore, aided by the AdS/CFT correspondence, a dual description of strongly coupled four-dimensional (4D) gauge theories can be obtained~\cite{Maldacena:1997re, ArkaniHamed:2000ds, Rattazzi:2000hs}. These fields are normally treated as test fields, where the corresponding energy density
is sufficiently small not to disturb the background geometry. Therefore, for a given geometry one performs a Kaluza-Klein (KK) decomposition. Solving the field equations in this geometry gives the bulk profile and the eigenmasses of the KK modes. The profile, in turn, determines the couplings of the KK modes.

Eventually, a complete model has to provide the dynamical elements that determine the bulk geometry, and the brane positions (when present). As an example, the 
Randall-Sundrum model \cite{Randall:1999ee}, does not solve the hierarchy problem until the exact location of the TeV brane is fully specified. Most of stabilization mechanisms, starting from  \cite{Goldberger:1999uk}, typically employ one scalar field, due to its simplicity.

The system of perturbations in the presence of a bulk scalar was studied, for example, in Ref.~\cite{Csaki:2000zn} (see also \cite{Lesgourgues:2003mi}). This analysis considered the properties of various excitations coupled to SM fields, which were assumed to be localized on the brane. Of particular relevance was the study of the radion, which is the lightest of these scalar perturbations. The analysis of \cite{Csaki:2000zn} was limited to a regime of small backreaction of the bulk scalars on the bulk geometry. Later, \cite{Kofman:2004tk} generalized this study to a bulk field with arbitrary bulk/brane potential.

However, there are systems in which more than one dynamical scalar field is relevant. For instance, in the soft-wall model of \cite{Batell:2008zm}, two bulk scalar fields are used to obtain a solution of the Einstein equations.
Furthermore in string theory, multiple scalar fields (such as the dilaton and 
moduli fields) are quite common. This motivates generalizing the study 
of \cite{Kofman:2004tk} to the case of $N$ bulk (real) scalar fields, which will be presented in this paper.
The generalization to $N$ bulk scalar fields has also been previously considered in 
Refs.~\cite{Toharia:2010ex,Aybat:2010sn,George:2011tn}. In our analysis 
we make no assumption on the form of the bulk and boundary potentials.
 
The perturbations are obtained by studying the linearized problem. It is useful to distinguish them as scalar/vector/tensor modes under Lorentz transformations of the ordinary 4D spacetime: we assume that the noncompact space has a Minkowski metric. As a consequence the scalar/vector/tensor modes are decoupled from each other at the linearized level and therefore can be studied separately. The mode count proceeds as follows. There are 15 perturbations in the five-dimensional (5D) metric, and $N$ perturbations from the bulk (real) scalar fields. Of these, $5$ are nondynamical and 5 more can be removed by a gauge choice\footnote{Gauge invariant variables can also be used. We choose the explicit gauge choice here, for brevity of exposition, but the two approaches are completely equivalent.}. This leaves $5+N$ modes of which, $5$ are in the tensor sector (transverse and traceless $h_{\mu \nu}$) and $N$ remain as scalar modes. There are no vector modes. 
Finally, there is a scalar mode for each brane (representing the fluctuation of its bulk position); however, we show that these modes are not excited as long as the branes have only a tension (namely, no other brane source with a different equation of state is dynamically relevant; this is  the usual assumption in brane compactifications) and so we will ignore them.

It is technically challenging to identify the $N$ scalar degrees of freedom because they can arise either from the components of the metric perturbations or the actual bulk scalar field perturbations. The number can be reduced to $N$ by using constraint equations to derive a manageable set of closed equations for 
the $N$ modes. This must be done for both the bulk and brane equations. The latter equations are boundary conditions and we will consider both cases either with or without branes at the boundaries (in the second case,  the boundary condition is typically given by the requirement of normalizability). We derive this system of equations, which for convenience is written as an eigenvalue problem, with the appropriate number of equations needed to obtain a solution. Our goal is to provide an explicit formulation of the eigenvalue problem that can be used to study any 5D model with an arbitrary number of bulk scalar fields.  By solving the eigenvalue problem the physical properties of the perturbations (masses and couplings) can be obtained. This can then be used to
check the stability of the model, so that if $m^2<0$ for some scalar modes, the background solution is unstable.

While the eigenvalue problem may be solved analytically for the simplest cases, 
in general a numerical method is needed to obtain a solution. We will employ the shooting method since the boundary equations at one boundary leave freedom for the choice of the mode functions, and of the mass eigenvalue. We show that, for $N$ scalars, the boundary conditions at one of the two boundaries
 leave $N$ unspecified quantities. The bulk equations are then used to evolve the solution to the other boundary, where there are precisely $N$ constraints that must be satisfied (given by the boundary condition at this boundary). If these constraints are satisfied, then a physical mode has been found.

As an example of the numerical method, we study the perturbation properties, and the stability problem, for the dynamical soft wall model of Ref.~\cite{Batell:2008zm}. The model is characterized by two scalar fields with a coupled potential term. It is interesting because it leads to a KK mass spectrum with linear Regge-like trajectories, similar to the hadron spectrum in QCD. We conduct a dense scan in parameter space and find approximately 100 modes with positive $m^2$, and no mode with negative $m^2$. Moreover, we find an interesting behavior of the radion mass. The mass is suppressed by a large logarithm, that in the dual CFT interpretation corresponds to how scale invariance is broken by quantum corrections. This is the same suppression present  in the Goldberger-Wise mechanism \cite{Goldberger:1999uk,Csaki:2000zn}.

The paper is organized as follows. In Section \ref{models} we introduce the class of  models that we are studying; we present the Lagrangian, the background solutions, and the most general set of perturbations. In the following three Sections we study the scalar, vector, and tensor perturbations, respectively. We identify the physical modes, and perform their Kaluza-Klein decomposition. We provide the explicit closed set of equations which can be solved to obtain the mass and the bulk profile of these modes through a boundary value problem. In Section \ref{eigen-problem} we formalize this boundary value problem for the two sectors (scalar and tensor) that contain physical perturbations. In Section \ref{soft} we study the perturbations of the model  in Ref.~\cite{Batell:2008zm} as 
an example of how to use our formalism to solve the eigenvalue problem. In Section \ref{conclusions} we briefly summarize our main findings. Some more technical steps are then given in the three Appendices.

\section{5D Models}
\label{models}
The goal of this work is to provide the tools for studying the perturbations, and the stability, of a wide class of models with one extra dimension and $N$ bulk scalar fields  $\left\{ \varphi_i \right\}$. Specifically, we consider models characterized by the action
\begin{eqnarray}
S &=& 2   \int  d^5 x \sqrt{-g} \left[ \frac{M^3}{2} \, R - \frac{1}{2} \left( \partial \varphi_i \right)^2 - V \left( \varphi_i \right) \right] \nonumber\\
&&- \sum_{\rm branes} \int d^4 x \sqrt{-\gamma} \left\{ 2 M^3 \left[ K \right]_J + U \left( \varphi_i \right) \right\}\,,
\label{action}
\end{eqnarray}
where $M$ is the 5D fundamental scale. The first line in (\ref{action}) contains the bulk terms, which are restricted to lie between two boundary branes, if they are both present, or else either between a boundary brane and infinity, or between $\pm \infty$. The overall factor of $2$ is adopted from  \cite{Kofman:2004tk}, where the bulk was assumed to be $Z_2$ symmetric across each brane, and  the symmetry was accounted for by restricting the bulk integral only to one side of each brane. This notation will also be used here, even in the cases where one or both branes are absent; it is trivial to reabsorb this factor away by a  rescaling of $M$, $\varphi_i$, and $V$. We assume that the scalars have a standard kinetic term in the bulk. Notice that the choice of sign for the kinetic term corresponds to $\eta_{\mu \nu} = {\rm diag } \left( - 1 ,\, 1 ,\, 1 ,\, 1 \right)$. 
If present, a bulk cosmological constant is included as a constant term in $V$. The second line in (\ref{action}) is instead the brane action; $\gamma$ denotes the induced metric on the brane, and the notation $\left[ K \right]_J$ with subscript $J$ denotes the jump of the quantity inside the square parenthesis across the brane, which in this case is the extrinsic curvature $K$.
$U$ denotes the potential of the scalars on the brane (which is a function of the value that the $\varphi_i$ fields have at the brane location), and, if present,  a brane tension is included as a constant term in $U$ (note that when both boundary terms are present, we do not require them to be equal). In addition 
we disregard the possibility of brane kinetic terms for the scalar fields.

From the action (\ref{action}) we obtain the Einstein equations in the bulk 
\begin{equation}
G_{AB} = \frac{T_{AB}}{ M^3}~,
\label{einstein-eq}
\end{equation}
where capital Latin indices run over all the coordinates and the energy momentum tensor is given by
\begin{equation}
T_{AB} =  \partial_A \varphi_i \partial_B \varphi_i +  g_{AB} \left[ - \frac{1}{2} \left( \partial \varphi_i  \right)^2 - V \left(  \varphi_i \right) \right]\,.
\end{equation}
We also obtain the $N$ bulk equations for the scalars
\begin{equation}
\frac{1}{\sqrt{-g}} \,  \partial_A \left( \sqrt{-g} \,  g^{AB} \partial_B \varphi_i \right) - \frac{\partial V}{\partial \varphi_i} = 0~.
\label{scalar-eq}
\end{equation}
Note that not all of the equations in (\ref{einstein-eq}) and (\ref{scalar-eq}) are independent, due to the Bianchi identities.

When a boundary brane is present, extremizing the action (\ref{action}) leads to the boundary conditions for the scalar fields
\begin{equation}
\left[ n \cdot \nabla \varphi_i \right]_J = \frac{\partial U}{\partial \varphi_i}~,
\label{jun-phi}
\end{equation}
as well as the so-called Israel conditions
\begin{equation}
\left[ {\hat K}_{\mu\nu} \right]_J = - \frac{S_{\mu\nu}}{M^3}~,
\label{israel}
\end{equation}
where ${\hat K}_{\mu\nu} = K_{\mu\nu} - K \, \gamma_{\mu\nu} $, and 
$S_{\mu\nu}$ denotes the stress energy tensor on the brane
\begin{equation}
S_{\mu\nu} = - U \, \gamma_{\mu\nu}~.
\label{Smunu}
\end{equation}
Note that Greek indices run over the usual $3+1$ dimensions only. The explicit definition and computation of the induced metric and the extrinsic curvature are  given in Appendix A.

\subsection{Background }

At the background level, we assume a factorizable geometry with 4D Minkowski 
slices:
\begin{equation}
d s^2 =  A^2 \left( z \right) \left[ \eta_{\mu \nu} d x^\mu d x^\nu + d z^2 \right]~.
\end{equation}
It follows that the background bulk scalars can have a nontrivial dependence on the extra coordinate only:
\begin{equation}
\varphi_i = \phi_i \left( z \right)~.
\end{equation}
The only nontrivial equations in (\ref{einstein-eq}) then arise from the diagonal $\mu \nu$, 
and $z$ components.
We write here one linear combination of these two equations, and the $zz$ component:
\begin{eqnarray}
\frac{A''}{A} & = & 2 \frac{A^{'2}}{A^2}  - \frac{\phi_i^{'2}}{3 M^3}~, \nonumber\\
6 M^3 \frac{A^{'2}}{A^2} & = & \frac{\phi_i^{' 2}}{2} - A^2 \, V~,
\label{ein-bck}
\end{eqnarray}
where prime $(')$ denotes differentiation with respect to $z$. The scalar equations 
(\ref{scalar-eq}) give instead
\begin{equation}
\phi_i'' + 3 \frac{A'}{A} \phi_i' - A^2 \, V_{,i} = 0~,
\label{sca-eq-bck}
\end{equation}
where $V_{,i} \equiv  \frac{\partial V}{\partial \phi_i}$. It is easy to check that the first equation 
of (\ref{ein-bck}) can be derived by combining (\ref{sca-eq-bck})  and the second equation of  (\ref{ein-bck}) (this redundancy is a consequence of a nontrivial component of the  Bianchi identity).

From the background  expressions of the induced metric and the extrinsic curvature given in 
Appendix A, the Israel conditions (\ref{israel}) have only a nontrivial part proportional to $\eta_{\mu \nu}$:
\begin{equation}
\frac{A'}{A^2} = \mp \frac{U}{6 M^3}~,
\label{bou-bck-ein}
\end{equation}
while the boundary conditions (\ref{jun-phi}) can be rewritten as
\begin{equation}
\frac{\phi_i'}{A} = \pm \frac{U_{,i}}{2}~,
\label{bou-bck-phi}
\end{equation}
where $U_{,i} \equiv \frac{\partial U}{\partial \phi_i}$. The upper (lower) sign at the right hand side of these two equations refers to a brane at the left (right) of the bulk interval. If one or both branes are absent, equations (\ref{bou-bck-ein}) and (\ref{bou-bck-phi}) can be replaced by different boundary conditions at spatial infinity along the bulk.

In the study of the perturbations, we often make use of the background equations written in this Subsection in order to simplify the linearized equations for the perturbations, without writing this explicitly each time.

\subsection{Perturbations}

It is convenient to characterize the perturbations according to how they transform under 4D Lorentz transformations. Due to the background symmetry, modes that transform differently under these transformations are decoupled at the linearized level, and can be studied separately in our analysis. We therefore have the following decomposition
\begin{eqnarray}
g_{\mu \nu} & = & A^2 \left( z \right) \Big[ \eta_{\mu \nu} \left( 1 + 2 \Psi \right) + 2 \partial_\mu \partial_\nu E \nonumber\\
&& \quad\quad\quad\quad
+ \partial_\mu E_\nu +  \partial_\nu E_\mu + h_{\mu \nu} \Big]~, \nonumber\\
g_{\mu 5} & = & A^2 \left( z \right) \left[  \partial_\mu B + B_\mu \right]~,  \nonumber\\
g_{5 5} & = & A^2 \left( z \right) \left[ 1 + 2 \, \Phi \right]~.
\label{metric-modes}
\end{eqnarray}
The modes $\Psi ,\, E ,\, B ,\, \Phi$ are scalar
(with respect to 4D Lorentz transformations); the modes $E_\mu ,\, B_\mu$ are vector (we impose that they are  transverse, $\partial_\mu E^\mu = \partial_\mu B^\mu = 0$), and  $h_{\mu \nu}$ is 
a tensor (imposed to be symmetric, transverse, and traceless, $\partial^\mu h_{\mu \nu} = h^\mu_\mu = 0$). There are also additional scalar modes:  the perturbations, $\delta \phi_i$ of the bulk scalars $ \varphi_i = \phi_i(z) + \delta \phi_i$, and the perturbations, $\zeta_j$ of the brane positions  
$z_{\rm j-th \; brane } = z_{\rm background \; position} +  \zeta_j ( x^\mu)$ ($j$ runs over the number of branes). All the perturbations are functions of both $x^\mu$ and $z$, except for $\zeta_j$ which are functions of $x^\mu$ only. Note that the decomposition (\ref{metric-modes}) becomes ambiguous for massless scalar KK modes, but can be studied using the light cone decomposition of~\cite{Kiritsis:2006ua}. We assume that there are no massless scalar modes in the cases of interest.

We need to fix the freedom of general coordinate transformations. Under the infinitesimal transformation
\begin{equation}
x^A \rightarrow x^A + \xi^A \;\;,\;\; \xi^A = \left( \partial^\mu \xi + \xi^\mu ,\, \xi^5 \right)~,
\label{change}
\end{equation}
(with $\partial_\mu \xi^\mu = 0$) the metric changes as
\begin{equation}
\delta g_{AB} \rightarrow \delta g_{AB} - g_{AB,C}^{(0)} \xi^C - g_{AC}^{(0)} \xi^C_{,B} - 
 g_{CB}^{(0)} \xi^C_A ~.
\end{equation}

We can use this relation to see how the various modes in (\ref{metric-modes}) transform. In particular we obtain:
\begin{eqnarray}
&&
E \rightarrow E - \xi \;\;,\;\; B \rightarrow B - \xi' - \xi^5 \;\;, \nonumber\\
&&
E_\mu \rightarrow E_\mu - \xi_\mu~,
\label{infinitesimal-metric-transf}
\end{eqnarray}
where we have only given the transformations relevant for the present discussion. This allows us to set $E_\mu =0$ in (\ref{metric-modes}) and removes the freedom of the transformations in (\ref{change}) characterized by 
$\xi_\mu$. Similarly we can also set $E=0$ in  (\ref{metric-modes}), which fixes 
$\xi$, and choosing $B =0$ then fixes $\xi^5$. Therefore we see that one can always choose the gauge $E_\mu = E = B = 0$; this completely fixes the 
freedom of the coordinate transformations (\ref{change}). This leaves the sets of scalar  $\left\{ \Psi ,\, \Phi ,\, \delta \phi_i ,\, \zeta_j \right\}$,  vector  $\left\{ B_\mu \right\}$ and tensor $\left\{ h_{\mu \nu} \right\}$ modes; these three systems are decoupled from each other at the linearized level, and we will study them separately in the following three sections.

\section{Scalar perturbations}
\label{scalar}

In this Section we write the linearized equations for the scalar perturbations. The main goal is to remove the nondynamical degrees of freedom. As we wrote in the last paragraph of the previous Section, we start from the system of perturbations $\left\{ \Psi ,\, \Phi ,\, \delta \phi_i ,\, \zeta_j \right\}$, where $i$ runs over the number of bulk scalars ($N$), while $j$ runs over the number of branes. 

We can immediately show that the brane displacements are decoupled, and do not introduce any instability. They only enter in the boundary conditions. Specifically, let us assume that a brane is present and consider the linearization of  (\ref{israel}) at that location (omitting the index $j$ on the displacement $\zeta_j$). Using the results in Appendix A, we obtain
\begin{eqnarray}
&&\left[ \left\{  A' \left( 1 - \Phi + 2 \Psi \right) + A \Psi' \right\} \eta_{\mu \nu} - A \zeta_{,\mu \nu} \right]_J \nonumber\\
&&\quad\quad=\frac{U}{M^3} A^3\left( 1 + 2 \Psi \right) \eta_{\mu \nu}~.
\label{israel-scalar}
\end{eqnarray}
(Note that $U$ also has a perturbation part). This equation has two tensorial structures that need to vanish independently. In particular, we find that $\left[ \zeta_{,\mu \nu} \right] _J= 0$. Using the $Z_2$ symmetry at the brane, and the fact that $\zeta$ is odd, this equation in turns gives 
\begin{equation}
\Box \zeta = 0~,
\end{equation}
where $\Box$ denotes the d'Alembertian operator in 4D Minkowski space.
As is well known for the single scalar case, the brane bending mode is not sourced by the scalar fields, and is just a decoupled massless mode in the current context. Therefore, in the following we simply disregard this brane mode(s).

We are left with the $N+2$ scalar perturbations  $\left\{ \Psi ,\, \Phi ,\, \delta \phi_i \right\}$. As we discussed in the Introduction, there are only $N$ physical perturbations in the scalar sector. In Subsection \ref{scalar-bulk} we present the linearized bulk equations for the scalars, and show how the two nondynamical modes can be eliminated from two constraint equations. We actually define $N$ scalar combinations $v_i$, that correspond to the canonical variables of the system. In the following three subsections we then compute the boundary conditions for these variables for different relevant cases.

\subsection{Bulk equations}
\label{scalar-bulk}

We start by considering the  scalar projection of the linearized Einstein equations (\ref{einstein-eq}) in the bulk. The $\mu \nu$ components read 
\begin{equation}
\left( \dots \right) \eta_{\mu \nu} - \left( \Phi + 2 \Psi \right)_{,\mu \nu} = 0~.
\label{con0}
\end{equation}
The off-diagonal part requires that
\begin{equation}
\Psi = - \frac{\Phi}{2}~.
\label{con1}
\end{equation}
From now on, we will enforce this constraint to eliminate $\Psi$. The $\mu5$ linearized equations then give
\begin{equation}
\frac{1}{2} \Phi' + \frac{ A'}{A} \Phi - \frac{1}{3 M^3} \phi_i' \delta \phi_i = 0~.
\label{con2}
\end{equation}
Eqs. (\ref{con1}) and (\ref{con2}) are the two constraints that allow the remaining  two nonphysical modes in the scalar sector to be eliminated after the gauge fixing.

Next, we introduce the $N$ combinations
\begin{equation}
v_i \equiv  - \sqrt{2} A^{3/2} \left( \delta \phi_i + \frac{A \phi_i'}{2 A'} \, \Phi \right)~,
\label{def-v}
\end{equation}
which generalize the canonical variable introduced in \cite{Kofman:2004tk} for the case of a single scalar field. These modes are the canonical variables of the system. After the conditions (\ref{con1}) and (\ref{con2}) are used to eliminate the nondynamical modes, one can show (see Appendix B) that the remaining bulk equations are equivalent to the following system of $N$ equations for the $N$ modes $v_i$:
\begin{eqnarray}
&&\Box v_i + v_i'' - \Bigg[ A^2 V_{,ij} + \frac{2 A^3}{3 M^3 A'} \left( V_{,i} \phi_j' +  V_{,j} \phi_i' \right)
 \nonumber\\
&& + \frac{2 \phi_i' \phi_j'}{3 M^3} \left( - 4 + \frac{A^2 \phi_k' \phi_k'}{3 M^3 A^{'2}} \right) 
\nonumber\\
&&+ \left( \frac{15}{4} \frac{A^{'2}}{A^2} - \frac{\phi_k' \phi_k'}{2 M^3} \right) \delta_{ij} \Bigg] v_j = 0~.
\label{eq-v}
\end{eqnarray}
This is the explicit set of equations for the dynamical scalar modes of the system.

\subsection{Boundary conditions for finite brane potential}
\label{generalU}

After disregarding $\zeta$, only the term proportional to $\eta_{\mu\nu}$ survives in the Israel conditions
(\ref{israel-scalar}). Using (\ref{con1}) and the $Z_2$ symmetry, we rewrite it as
\begin{equation}
24 \frac{A'}{A^2} \left(  \Phi + \frac{A \Phi'}{2 A'} \right)   \Bigg\vert_{\rm brane}  = \pm \frac{4}{M^3} \, U_{,k} \, \delta \phi_k~,
\end{equation}
where the left hand side is evaluated in the bulk immediately next to the brane, and the upper (lower) sign at the right hand side refers to a brane at the left (right) of the bulk interval. This equation does not provide any additional information with respect to the bulk equations. We can indeed rewrite the right hand side in terms of bulk quantities using 
eq.(\ref{bou-bck-phi}); the resulting equation is simply the constraint equation (\ref{con2}) at the brane location. The linearization of (\ref{jun-phi}) is instead nontrivial:
\begin{equation}
\delta \phi_i' -  \phi_i' \Phi  \Big\vert_{\rm brane} = \pm \frac{A}{2}  U_{,ij} \delta \phi_j~,
\label{jun2}
\end{equation}
where again  the left hand side is evaluated in the bulk immediately next to the brane.

The role of the boundary conditions is to complement  the bulk equations (\ref{eq-v}) and form an eigenvalue problem that can be immediately solved to obtain the physical scalar excitations of the system. By looking at eqs. (\ref{eq-v}), we see that the most useful form in which these equations can be written is $v_i' = \sum_j c_{ij} v_j$, where the coefficients $c_{ij}$ depend on background quantities. Considerable algebra is required to obtain this equation starting from (\ref{jun2}). The final result is
\begin{eqnarray}
&&v_i'\Big\vert_{\rm brane}  =    \frac{3 A'}{2 A} v_i + \left( \frac{A \phi_i' \, \phi_j'}{3 M^3 A'} \pm \frac{A}{2} U_{,ij} \right) v_j 
\nonumber\\
&& 
+  \frac{    4 \frac{A'}{A} \phi_k' v_k - A^2 V_{,k} v_k \pm \frac{A}{2} U_{,rs} \phi_r' v_s  }{ \frac{3 M^3 }{\sqrt{2}} \Box - 2 \sqrt{2}  \phi_p' \phi_p' + \frac{A^3 \phi_p' V_{,p}}{\sqrt{2} A'} \mp  \frac{A^2}{2 \sqrt{2} A'} U_{,pq} \, \phi_p' \phi_q' } 
\nonumber\\
&&
\quad\quad\quad
\times \left( 2 \sqrt{2}  \phi_i' - \frac{A^3 V_{,i}}{\sqrt{2} A'} 
 \pm \frac{A^2 \,  U_{,ij} \, \phi_j' 
}{2 \sqrt{2} A'}\right)~,
\label{junction}
\end{eqnarray}
which are indeed equivalent to (\ref{jun2}) (we show this  in Appendix C). We stress that these equations are valid at the brane location; brane quantities are evaluated immediately next to the brane, and whenever an upper/lower sign appears on the right hand side, it refers to a brane at the left/right of the bulk interval, respectively. Finally, let us clarify the role of the  $\Box$ operator on the right hand side. The bulk equations (\ref{eq-v}) allow for a factorizable solution
\begin{eqnarray}
v_i \left( x^\lambda ,\, z \right) & = & \sum_n  {\tilde v}_i^{(n)} \left( z \right)  \, Q^{(n)} \left( x^\lambda \right)~, 
\label{KK5}
\end{eqnarray}
with
\begin{eqnarray}
\Box  Q^{(n)} & = & m_n^2  Q^{(n)}~.
\label{KK4}
\end{eqnarray}
Eq. (\ref{KK5})  is the decomposition in Kaluza-Klein modes; each mode is characterized by a wave function $ {\tilde v}_i^{(n)} $  in the bulk, as well as a 4D (quantum) field $ Q^{(n)} $. Equations (\ref{eq-v}) and (\ref{junction}), with the substitution $v_i \rightarrow {\tilde v}_i^{(n)}$ and $\Box \rightarrow m_n^2$, provide the complete eigenvalue problem to determine the eigenmasses and the bulk profiles of the scalar modes of the system. It is clear from the form of the equations that this problem is well posed, and can be uniquely solved. We discuss  this in detail in Section \ref{eigen-problem}, and provide an explicit example in Section \ref{soft}.

\subsection{Boundary conditions for (infinitely) stiff  brane potential}

\label{stiff}

One can obtain a simpler set of boundary conditions than (\ref{junction}) in the limit of infinitely stiff brane potentials. Let us Taylor expand the brane potential for small fluctuations $\delta \phi_i$ around the background values $\phi_i$:
\begin{equation}
U \left( \varphi_i =  \phi_i + \delta \phi_i \right) = U  + U_{,i} \, \delta \phi_i + \frac{1}{2} U_{,ij} \delta \phi_i \, \delta \phi_j + \dots~,
\label{taylor-U}
\end{equation}
where the potential and its derivatives on the right hand side are evaluated at the background solution $\phi_i$. As can be seen from (\ref{bou-bck-ein}) and (\ref{bou-bck-phi}), only the expectation values of the brane potential $U$ and its first derivatives $U_{,i}$ are relevant at the background level. The brane ``masses'' $m_{ij}^2 \equiv U_{,ij}$ only enter in the boundary conditions for the linear perturbations; higher-order terms in (\ref{taylor-U}) are instead relevant only beyond the linearized level, and can be disregarded in our study. The stiff potential limit is the limit for which the ``masses'' $\sqrt{m_{ij}^2}$ are much greater than any other mass scale in the problem. In the original Goldberger-Wise \cite{Goldberger:1999uk} stabilization mechanism, this is the limit of large $\lambda_{h,v}$; it is also explicitly noted there, that the equations considerably simplify in this limit. This limit can always be imposed by simply adding a quadratic potential term $\Delta U = \lambda_{ij} \left( \varphi_i - \phi_i \right) \left( \varphi_j - \phi_j \right)$ centered on the background solution, and then taking the limit of large $\lambda_{ij}$. Adding this term does not modify the background solution. 

The simplification occurs because in this limit $\delta \phi_i \rightarrow 0$ at the brane location; this can be seen from eqs.~(\ref{jun2}). More precisely  the fluctuations  $\delta \phi_i$ on the brane ``adjust themselves'' to the ${\cal O } \left( U_{,ij}^{-1} \right)$ level required to satisfy  (\ref{jun2}). We actually do not need the explicit solutions for the scalar fluctuations. If we consider the equations in the original set of variables $\left\{ \delta \phi_i ,\, \Phi \right\}$ we see that the boundary conditions (\ref{jun2}) are the only terms in which the second derivatives  $U_{,ij}$ are present. These equations can be solved for sufficiently small $\delta \phi_{i,{\rm brane}}$, but then, once they are satisfied, the only role that these equations play in the stiff limit is to impose that $\delta \phi_{i,{\rm brane}}$ at the boundary can be set to zero in all the other equations of the system. The situation is completely analogous to the scattering of light massive objects against an infinitely heavy object. An object of large mass $M$ acquires an infinitesimally small, ${\cal O } \left( 1 / M \right)$, velocity in the scattering.  For $M \rightarrow \infty$ we simply disregard the motion of the heavy object, and the value of $M$ drops from the problem; the role of the heavy object is to ensure that the momentum conservation equation is satisfied, but then this momentum conservation equation plays no role in the dynamics of the light mass(es) participating in the scattering. Eqs. (\ref{jun2}) are analogous to the momentum conservation equations, $U_{,ij}$ is the analog of the mass $M$, and $\delta \phi_{i,{\rm brane}}$  are analogous to the velocity acquired by the heavy object. 

In this limit, eq. (\ref{def-v}) then imposes the condition 
\begin{equation}
v_{i,{\rm brane}} = {\cal N} \, \phi_{i,{\rm brane}}'~,
\label{v-bc-stiff}
\end{equation}
where the proportionality constant ${\cal N}$ is the same for all modes. Eq.~(\ref{v-bc-stiff}) provides a system of  $N-1$ independent  boundary conditions in the eigenvalue problem. The reason why the number is  $N-1$ (rather than $N$) is because the overall normalization of a mode - which is proportional to ${\cal N}$ -  cannot be specified by the linearized problem we are solving (if one multiplies the solutions of a linear system by a common factor, one still has a solution). Fortunately, we do not need to know ${\cal N}$  if we are only interested in solving the linearized problem for the eigenmasses of the modes,  and therefore we can simply set ${\cal N}$ to any convenient nonvanishing value 
(see Section~\ref{eigen-problem}).~\footnote{The normalization of the modes, or, equivalently, the value of ${\cal N}$ in eqs. (\ref{v-bc-stiff}),  is needed if one instead wants to compute the couplings of the perturbations to each other, or to other fields (because the couplings are determined by the actual value of the wave function). The normalization is obtained by canonically normalizing the modes in the quadratic action of the perturbations, see Subsection  \ref{kin-action}.}

The  ``missing'' boundary condition, in addition to the  $N-1$ independent boundary conditions (\ref{v-bc-stiff}),  is obtained by evaluating the constraint equation (\ref{con2}) at the brane
location, for $\delta \phi_{i,{\rm brane}} = 0$ :
\begin{equation}
\Phi' + \frac{2 A'}{A} \, \Phi \; \big\vert_{\rm brane} = 0~.
\label{Phi-bc-stiff}
\end{equation}
We need to rewrite this equation in terms of $v_i'$ and ${\cal N}$. Using (\ref{v-bc-stiff}) and  $\delta \phi_{i,{\rm brane}} = 0$ in eqs. (\ref{def-v}),  we obtain 
\begin{equation}
\Phi  \big\vert_{\rm brane} = - \frac{\sqrt{2} \, A'}{A^{5/2}} \, {\cal N}~.
\label{Phi-stiff}
\end{equation}
To obtain the desired expression for $\Phi'|_{\rm brane}$, we then differentiate with respect to $z$ the second of the bulk equations (\ref{bulk-sca-i55}).~\footnote{Note that eq.~(\ref{Phi-stiff}) is a boundary condition, and so we cannot obtain an expression for $\Phi'|_{\rm brane}  $ by simply differentiating it.} Using (\ref{Phi-stiff}) in the resulting expression, we obtain a relation for $\Phi'|_{\rm brane}$ in terms of  $v'_{i,{\rm brane}}$ and  ${\cal N}$. Using this relation, and eq.~(\ref{Phi-stiff}), we can then cast  the condition (\ref{Phi-bc-stiff}) into the desired form. 
In terms of the bulk wave functions, the condition reads:
\begin{eqnarray}
 \frac{1}{\cal N} \phi_i' {\tilde v}_i^{'(n)}  \Big\vert_{\rm brane}\!\! &= & A^2 \phi_i' V_{,i} - 3 M^3 \frac{A'}{A} m_n^2 \nonumber\\
  && - \frac{5}{2} \frac{A'}{A} \phi_i' \phi_i' + \frac{A}{3 M^3 A'} \left( \phi_i' \phi_i' \right)^2~.
\label{bc-stiff}
\end{eqnarray}

In the stiff brane limit the $N$ equations (\ref{v-bc-stiff}) and (\ref{bc-stiff}) replace the conditions (\ref{junction}).

\subsection{Boundary conditions without a boundary  brane}

\label{nobrane}

Next we comment on the possibility that one or both boundary branes are absent. Assume that the bulk coordinate extends to infinity in that particular direction(s). In this case, the boundary conditions for the perturbations can be dictated by the specific problem under consideration. A typical requirement is the one of normalizability. The set of differential equations (\ref{eq-v}) has $2 N$ solutions. In the example that we study in Section \ref{soft}, one finds that half of these solutions exponentially grow at $z \rightarrow \infty$, while the remaining half exponentially decrease. Therefore, eliminating the exponentially growing solutions provides precisely $N$ boundary conditions, as it was the case for the boundary conditions enforced by a boundary brane. Other models are characterized by a horizon at some bulk position, and one then typically requires that the solutions should be purely infalling modes at the horizon. Also this requirement corresponds to $N$ boundary conditions.

\subsection{Quadratic scalar action, and normalization of the scalar modes}
\label{kin-action}

To properly normalize the scalar modes, we compute the kinetic term of their quadratic action, obtained by expanding the starting action (\ref{action}) at second order in these perturbations. We find
\begin{eqnarray}
S_{2,{\rm kin}} \!\!  &=& \!\!  \int d^5 x A^3  \eta^{\mu \nu} \left[ 6 M^3 \partial_\mu \Psi \partial_\nu  \left(\Psi  +  
\Phi \right)  -  \partial_\mu \delta \phi_i \partial_\nu \delta \phi_i  \right] \nonumber\\
 & = & \!\! \int d^5 x  \left[ \frac{1}{2} v_i \Box v_i + \partial_y \left(
\frac{3 M^3 A^4}{4 A'} \eta^{\mu \nu} \partial_\mu \Phi \partial_\nu \Phi \right) \right] \;.\nonumber\\
\label{kin2}
\end{eqnarray}
If we replace $\Psi$ in the first line of this expression through the constraint equation (\ref{con1}), 
$\Psi = - \frac{\Phi}{2}$, we can immediately see that the kinetic term is manifestly positive (recall that $\eta^{00} = - 1$), which ensures that the scalar system has no ghosts. The second line of (\ref{kin2}) has been obtained  following the steps outlined  before eq. (19) of  \cite{Kofman:2004tk}, where an analogous computation was performed for the case of a single scalar field. By decomposing $v_i$ as in (\ref{KK5}), and, analogously, $\Phi = \sum_n \Phi^{(n)} \left( z \right) Q^{(n)} \left( x \right)$, we arrive at
\begin{equation}
S_{2,{\rm kin}}  = \sum_{m,n} C_{mn} \int d^4 x \,  Q^{(m)} \Box Q^{(n)}\,,
\end{equation}
where
\begin{equation}
C_{mn}  \equiv  \frac{1}{2} \,  \int d z \;  v_i^{(m)} \, v_i^{(n)}  + \frac{3 M^3 A^4}{4 A'} \, \Phi^{(m)} \, \Phi^{(n)}  \Big\vert_{z_{\rm min}}^{z_{\rm max}}\,.
\label{Cmn}
\end{equation}
In evaluating the boundary term, one can make use of eq. (\ref{Phi-v-vp}) to express  $\Phi^{(n)}$ in terms  
of $v_i^{(n)}$ and  $v_i^{\prime(n)}$. 

We note that the result (\ref{Cmn}) is the most immediate generalization to $N$ fields of the expression (26) obtained in  \cite{Kofman:2004tk} for the single field case (we note that the sign of the boundary term in the intermediate expression in eq. (26) of   \cite{Kofman:2004tk} is incorrect). Hermiticity of $S_2$ ensures that eigenmodes with different mass are orthogonal, so $C_{mn} \propto \delta_{mn}$. Imposing 
 $C_{nn} = \frac{1}{2}$, we then recover a diagonal and canonically normalized kinetic term
 \begin{equation}
S_{2,{\rm kin}}  = \sum_{n} \frac{1}{2}  \int d^4 x \,  Q^{(n)} \Box Q^{(n)}\,.
\end{equation}

If the linearized equations can be solved analytically, one can leave the normalization of the modes (specifically, the quantity ${\cal N}$, in the case of a stiff boundary potential), and then fix it through $C_{nn} = \frac{1}{2}$. If the equations can be only integrated numerically, one needs to set a provisory value for the normalization by (arbitrarily) fixing the value of one of the wave functions at one boundary (for instance, in the example that we study in Section \ref{soft} we set $v_2^{(n)} = 1$ at the UV boundary); after performing the numerical integration, one can then insert the solutions in  (\ref{Cmn}) and obtain the provisory result $C_{nn,{\rm prov}}$. The rescaling $v_i^{(n)} \rightarrow \frac{1}{\sqrt{2 \,  C_{nn,{\rm prov}}}} \, v_i^{(n)}$ provides the correctly normalized modes.

\section{Vector perturbations}

For the vector modes, the $5\mu$ and $\mu\nu$ components of the linearized Einstein equations (\ref{einstein-eq}) in the bulk give, respectively,
\begin{eqnarray}
\Box B_\mu &=& 0~, \nonumber\\
\left[ B_\mu' +  3 \frac{A'}{A} B_\mu  \right]_{,\nu} &=& 0~,
\end{eqnarray}
while the $55$ component trivially vanishes. Moreover, the linearization of equations (\ref{scalar-eq}) has no contributions from the vector modes. The bulk equations therefore enforce $B_\mu \left( x^\lambda , \, z \right) = b_\mu \left( x^\lambda \right)   A^{-3} \left( z \right)$, with $\Box b_\mu = 0$. This immediately implies that there are no Kaluza-Klein modes in the vector sector, except for a zero mode.

If branes are present, it is also immediate to see that the linearization of the boundary conditions for the scalar fields, eqs.~(\ref{jun-phi}) have no contributions from the vector perturbations. The linearization of the Israel junction conditions, eqs.~(\ref{israel}), is instead nontrivial. We see from (\ref{Smunu}) and  (\ref{gamma-sol}) that the brane stress-energy tensor is $B_\mu-$independent.  Using  (\ref{K-sol}), we therefore have
\begin{equation}
\left[ B_\mu \right] _J= 0~.
\end{equation}
Under the assumption of a $Z_2$ symmetry,  $B_\mu$ needs to be odd across the brane,\footnote{The simplest way to see this is to consider the gauge invariant combination out of the two vector perturbations appearing in the metric (\ref{metric-modes}). As shown in (\ref{infinitesimal-metric-transf}), $E_\mu \rightarrow E_\mu - \xi_\mu$ under the infinitesimal coordinate transformation (\ref{change}). One can also show that, under this transformation,  $B_\mu \rightarrow B_\mu - \xi_\mu'$. Therefore, the gauge invariant combination is $B_\mu - E_\mu'$. As $E_\mu'$ is odd across the brane, this implies that $B_\mu$ is also odd.} which then enforces $B_\mu = 0$. 
The absence of vector modes agrees with the counting of the number of physical degrees of freedom given in the Introduction.

\section{Tensor perturbations}

The  $\mu\nu$ components of the linearized Einstein equations (\ref{einstein-eq}) in the bulk give
\begin{equation}
\Box h_{\mu \nu} + h_{\mu \nu}'' + 3 \frac{A'}{A} \, h_{\mu \nu}' = 0~,
\label{eq-tensor}
\end{equation}
while the other components trivially vanish. Moreover, the linearization of equations (\ref{scalar-eq}) has no contributions from the tensor modes. 

If branes are present, we again find that the linearization of the boundary conditions for the scalar fields, eqs. (\ref{jun-phi}) have no contributions from the tensor perturbations. The linearization of the Israel junction conditions, eqs. (\ref{israel}) instead gives
\begin{equation}
\left[ h_{\mu \nu}' \right]_J = 0~.
\label{israel-tensor}
\end{equation}

We note that the tensor modes only ``respond'' to the background geometry, and not to the details of the sources. Therefore, these results are similar to those already derived for the case of a single scalar in the bulk (see, for example,  \cite{Frolov:2002qm}). In Ref. \cite{Frolov:2002qm} the  Kaluza-Klein eigenmasses were shown to be nonegative, so that there is no   instability in the   tensor sector. For completeness, we summarize this computation here:

One starts by decomposing the 5D tensor field as
\begin{eqnarray}
h_{\mu \nu} \left( x^\lambda ,\, z \right) & = & A^{-3/2} \left( z \right)  \, \sum_n {\tilde h}_n \left( z \right) \, Q_{\mu \nu}^{(n)} \left( x^\lambda \right), \nonumber\\
\Box  Q_{\mu \nu}^{(n)} & = & m_n^2 \,  Q_{\mu \nu}^{(n)}~,
\label{deco-tensor}
\end{eqnarray}
where ${\tilde h}_n$ satisfies the Schroedinger-like equation
\begin{equation}
{\tilde h}_n'' + \left[ m_n^2 - \frac{3 A''}{2 A} - \frac{3 A^{'2}}{4 A^2} \right] {\tilde h}_n = 0~.
\label{sch-tensor}
\end{equation}
One can further define the operators~\footnote{Notice that  \cite{Frolov:2002qm} 
defines ${\cal D}_\pm$ in the opposite way, since their $\Omega$ is $1/A$.}
\begin{equation}
{\cal D}_\pm \equiv \partial_z \pm \frac{3 A'}{2 A}~,
\end{equation}
so that the bulk and brane equations become, respectively,
\begin{eqnarray}
&& - {\cal D}_+ \, {\cal D}_- \, {\tilde h}_n = m^2 \, {\tilde h}_n~, \nonumber\\
&& \left[ {\cal D}_- \, {\tilde h}_n \right]_J = 0 \;\;\Rightarrow\;\; \left(  {\cal D}_- \, {\tilde h}_n \right) \Big\vert_{\rm brane} = 0~,
\label{dpdm}
\end{eqnarray}
where in the last step we have used the $Z_2$ symmetry across the brane.

If the first equation of (\ref{dpdm}) is multiplied by ${\tilde h}_n^*$ from the left and integrated over the bulk coordinate $z$, the left hand side can be then integrated by parts. The resulting boundary term then vanishes because of the second equation of (\ref{dpdm}) and leads to the condition \cite{Frolov:2002qm} 
\begin{equation}
m_n^2 = \frac{\int d z \vert {\cal D}_- \, {\tilde h}_n \vert^2}{\int d z \vert {\tilde h}_n \vert^2} \, \geq 0~.
\label{general-m-h}
\end{equation}

Using Eqs. (\ref{dpdm}) we can also immediately determine the existence of a tensor massless mode, characterized by the bulk profile
\begin{equation}
{\cal D}_- \, {\tilde h}_0 = 0 \;\;\Rightarrow\;\; {\tilde h}_0 \propto A^{3/2}~.
\end{equation}
From (\ref{deco-tensor}) and the metric decomposition (\ref{metric-modes}), we recover the well-known fact that the massless tensor mode has an identical bulk profile as the background geometry. Incidentally there is no issue with using the decomposition (\ref{metric-modes}) for the massless tensor mode because the tensor equation of motion does not change~\cite{Kiritsis:2006ua}.

\section{Eigenvalue problem}

\label{eigen-problem}

In the three previous Sections we have obtained the canonical modes in both the scalar and tensor sector, while we have shown that there are no physical vector modes. We have  decomposed the canonical perturbations in a  Kaluza-Klein sum, and obtained the equations satisfied by  the wavefunction of the modes. In the two following Subsections we outline the eigenvalue problem that can be solved to obtain the properties of the physical modes.

\subsection{Scalar sector}

When $N$ bulk scalars are present,  the scalar sector of the perturbations is characterized by the  $N$ physical 5D perturbations $v_i$, defined in (\ref{def-v}). The wave functions of the corresponding Kaluza-Klein modes satisfy the $N$ second order differential  equations (\ref{eq-v}). Each KK mode is therefore characterized by $2 N + 1$ parameters (the mass $m_n$, and the $2N$ values required to specify the Cauchy problem), so that the bulk differential equations need to be supplemented by $2 N + 1$ conditions. Each boundary brane enforces $N$ conditions, given by eq. (\ref{junction}) in the case of finite brane potentials, and by eqs. (\ref{v-bc-stiff}) and (\ref{bc-stiff}) in the case of infinitely stiff brane potentials. We discussed in Subsection \ref{nobrane} the typical boundary conditions that can be imposed in the case that one or both branes are absent (in general, we expect $N$ conditions per boundary). One additional  condition is obtained by fixing the overall normalization of the modes. 
One can typically start by fixing a provisory (and, generally, incorrect) normalization;  for instance, one can require that one of the wave functions evaluates to $1$ at one boundary. This, together with the $2 N$ conditions coming from the two boundaries, allows the system of linearized equations to be completely solved. In this way, one obtains the masses of the physical modes, and the wavefunctions, up to an overall, yet to be specified, normalization. We stress that the overall normalization of the solutions is an irrelevant quantity in a linearized system of equations (if we rescale all the modes of a solution by a common factor, we still have a solution). Therefore, the system of equations can be solved, and gives the correct values of the eigenmasses, for any arbitrary overall normalization. Still, the overall normalization is crucial to determine the couplings of the perturbations, (since the couplings  are determined by the values of the wavefunctions); Subsection  \ref{kin-action} explains how to rescale the solutions, so as to obtain the correct normalization,  once the linearized system has been solved.

In many cases, the bulk equations cannot be solved analytically; we expect this to be the norm in the scalar sector, where the perturbations are coupled in the equations (this happens even if the bulk potential is a sum of separate terms, due to the mixing of the scalar field perturbations with the metric perturbations). In this case, the eigenvalue problem needs to be solved with a shooting method; one fixes half of the $2N$ parameters that are necessary to determine the bulk evolution, by enforcing the $N$ conditions at one of the two boundaries. Next one guesses the remaining $N$ parameters, and solves the bulk equations which enable the wave functions to be evaluated on the other boundary.  If the resulting solutions happen to satisfy the boundary conditions also there, then one has obtained a physical mode of the system. Typically, the initial guess is not correct, and one needs to employ some numerical scheme to obtain the solutions. For example, one can compute by how much the wavefunctions evaluated at the second brane differ from the expected boundary conditions, as a function of the initial guesses. An $N$-dimensional Newton's method can then be implemented to find the zeros of this function. We perform this algorithm in the example studied in the next Section.

\subsection{Tensor sector}

The tensor sector  is significantly simpler than the scalar sector.  The wave functions ${\tilde h }_n$, defined in equation (\ref{deco-tensor}), satisfy the second order differential equation (\ref{sch-tensor}) in the bulk. Each solution is in principle characterized by three parameters: the mass $m_n$ of the eigenmode, and two integration constants $ C_{1n} ,\, C_{2n}$.   However, as for the scalar case,   only the ratio of these two constants, and not the overall normalization of the solution, can be determined from the linearized problem.  Therefore to just obtain the eigenmasses and 
bulk profiles we can simply fix an arbitrary normalization by requiring that ${\tilde h}_n$ acquires a nonzero (but arbitrary) value at a given position (typically, at one boundary brane). This gives one condition. The other two conditions are enforced by the boundary branes (each brane enforces one condition, given by the last expression in (\ref{dpdm})), or if a brane is  absent, by the requirement that the wavefunction  is normalizable. These three conditions then allow $m_n,  C_{1n} ,\,$ and $ C_{2n}$ to be determined with the bulk profiles known up to an overall normalization. The correct normalization can be then obtained from the quadratic action of the tensor modes, analogously to what we dicussed for the scalar sector.

\section{An explicit example: the dynamical soft wall}
\label{soft}

As an explicit example of the general method that we have outlined above, we will 
consider the dynamical soft-wall solution found in Ref. \cite{Batell:2008zm}. In addition 
to the metric, this solution involves two bulk scalar fields. It provides a dynamical
realization of the holographic soft-wall model for QCD \cite{Karch:2006pv, Karch:2010eg}, 
as well as applications to the electroweak sector of the Standard Model~\cite{Batell:2008me}.

\subsection{The 5D model}

We review the soft wall  background solution~\cite{Batell:2008zm} 
(with the only difference that we use our convention for $M^3$, that corresponds to $M^3/2$ in \cite{Batell:2008zm}). The model is characterized by the two fields $\phi$ and $T$, with the bulk potential
\begin{eqnarray}
V & = & \frac{1}{8}\nu^2 k^2 T^2 \, {\rm e}^{ \frac{ \nu T^2}{6 M^3 \left( 1 + \nu \right) } } + \frac{1}{2}\nu^2 \, k^2 \phi^2 \, {\rm e}^{ \frac{ 2 }{ \sqrt{ 3 } }  \, \frac{ \phi }{ M^{ 3/ 2 }  } } \nonumber\\
& &  - 12 k^2 \Bigg[ \left( 1 + \nu \right) \, \frac{M^{3/2}}{\sqrt{2}} \, {\rm e}^{\frac {\nu T^2}{12 M^3 \left( 1 + \nu \right)} }
\nonumber\\
&&\quad\quad\quad\quad
- \nu \left( \frac{M^{3/2}}{\sqrt{2}} - \frac{\phi}{\sqrt{6}} \right) {\rm e}^{\frac{1}{\sqrt{3}}  \frac{\phi}{M^{3/2}} } \Bigg]^2~.
\end{eqnarray}
One obtains the background solution
\begin{eqnarray}
A \left( z \right) & = & \frac{{\rm e}^{-\frac{2}{3} \left( \mu z \right)^\nu}}{k \, z}~, \nonumber\\
\phi \left( z \right) & = & \frac{2}{\sqrt{3}} M^{3/2} \left( \mu z \right)^\nu~, \nonumber\\
T \left( z \right) & = & - 2 \sqrt{2} \sqrt{1+1/\nu} M^{3/2} \left( \mu z \right)^{\nu/2}~.
\label{soft-sol}
\end{eqnarray}
The parameter $\nu$ is a dimensionless constant, while $\mu$ is the soft-wall mass scale.
As the bulk volume diverges at $z \rightarrow 0$, a UV brane is placed at $z_0 = 1 / k$. The potential $U$ on this brane is chosen so that the solutions (\ref{soft-sol}) satisfy the boundary conditions there: specifically, the  background values of $U$ and $U_{,i}$ are determined from (\ref{bou-bck-ein}) and (\ref{bou-bck-phi}). We then assume that the brane potential contains large quadratic terms (\ref{taylor-U}), so that the boundary conditions for the perturbations can be given in the infinitely stiff limit of Subsection \ref{stiff}.

The scalar field configuration provides a finite bulk geometry in the limit $z \rightarrow \infty$, so that there is no need to include a boundary IR brane at large $z$ (in fact this is the reason for why it is a ``soft wall'', as opposed to a sharp brane or hard-wall cut-off). As we will see in the next two Subsections, the requirement that the perturbations are normalizable there provides sufficient boundary conditions to fully determine them. We focus our study of the perturbations of this model to the choice $\nu=2$, as this gives rise to a linear Regge-like mass spectrum, $m_n^2 \propto n$, for the KK modes, which is similar to that encountered in the hadron spectrum of QCD.

\subsection{Tensor modes}

The tensor modes for the model were already studied in \cite{Batell:2008me}. We summarize these results here for completeness, and to provide an example of an eigenvalue problem that can be solved analytically (and that is technically simpler than the problem for the scalar modes studied in the next Subsection). 

It is  convenient to use the  variable $h_{\mu \nu} = \delta g_{\mu \nu}^{TT} / A^2 \left( z \right)$ (where TT denotes the transverse-traceless component) for the eigenvalue problem. The Kaluza-Klein decomposition is analogous to (\ref{deco-tensor}), and we denote by $h_n$ the wave function of the $n-$th mode ($h_n = A^{-3/2} \, {\tilde h}_n$, where $ {\tilde h}_n $ is introduced in  (\ref{deco-tensor})). In the background (\ref{soft-sol}), the bulk equation  eq. (\ref{eq-tensor}) becomes
\begin{equation}
\frac{d^2  h_n}{d {\tilde z}^2} - \left( 4 \,  {\tilde z} + \frac{3}{ {\tilde z}} \right)  \frac{d  h_n}{d {\tilde z}} 
+ {\widetilde m}_n^2 \, h_n = 0~,
\label{eq-h-z}
\end{equation}
where we have introduced the dimensionless quantities  ${\tilde z} \equiv \mu \, z$ and ${\widetilde m} \equiv m / \mu $. The boundary condition at the UV brane then has the form
\begin{equation}
\frac{d h_n}{d {\tilde z}} \Big\vert_{{\tilde z} = {\tilde \mu}} = 0~,
\label{eq-h-bound}
\end{equation}
(where ${\tilde \mu } \equiv \mu / k$ corresponds to the location of the UV brane) 
while the normalizability requirement at ${\tilde z} \rightarrow \infty$ translates into the requirement that the solution decreases sufficiently fast at large ${\tilde z}$. More precisely, since the wave function of the canonically normalized mode is ${\tilde h}_n = A^{3/2} \, h_n$, we require that $ \int d z A^3 \, h_n^2 < \infty$.

The normalizable solution  of  Eq. (\ref{eq-h-z})  is (up to a normalization constant)  the Kummer's confluent hypergeometric function
\begin{equation}
h_n =  U \left( - \frac{{\widetilde m}_n^2}{8} ,\, - 1 ,\, 2 {\tilde z}^2 \right)~. 
\label{h-sol}
\end{equation}
The eigenmasses are obtained by imposing (\ref{eq-h-bound}). In the limit  ${\tilde \mu} = \mu / k \ll 1$, we can expand (\ref{h-sol}) at small ${\tilde z}$, and obtain
\begin{equation}
h_n =  {\rm const.} + \frac{16 {\tilde z}^2}{\left( - 8 + {\tilde m_n}^2 \right) \, \Gamma \left( - \frac{{\widetilde m}_n^2}{8} \right)} + {\cal O } \left( {\tilde z}^4 \right)~.
\end{equation}
Eq. (\ref{eq-h-bound}) is approximately satisfied at the poles of the gamma function, namely for 
\begin{equation}
{\widetilde m}_n \simeq 2 \sqrt{2 \, n} \;,\;\;\;\;\; n = 0 ,\, 2 ,\, 3 ,\, \dots~,
\end{equation}
(we verified numerically that there is indeed no physical mode corresponding to $n=1$). We note that ${\widetilde m}_n^2 \geq 0$ in agreement with the general result (\ref{general-m-h}).

\subsection{Scalar modes}

For the background solution (\ref{soft-sol}), and in terms of the dimensionless quantities ${\tilde z} ,\, {\widetilde m}_n$ defined in the previous Subsection, the bulk equations (\ref{eq-v}) become
\begin{eqnarray}
v_1'' \!\!\!\!  & = & \!\!\!\!  \left\{ - {\widetilde m}^2 + \frac{-9+8 {\tilde  z}^2 \left[ - 9 + 16 {\tilde  z}^2 \left( - 3 + {\tilde  z}^2 + 2 {\tilde  z}^4 \right) \right]}{4 {\tilde  z}^2 \left( 3 + 4 {\tilde  z}^2 \right)^2} \right\} v_1 \nonumber\\
& &  - \frac{16 {\tilde  z } \left( - 9 + 24 {\tilde  z}^2 + 16 {\tilde  z}^4 \right)}{3 \left( 3 + 4 {\tilde  z}^2 \right)^2} v_2~, \nonumber\\
v_2'' \!\!\!\! & = & \!\!\!\! \left\{ - {\widetilde m}^2 + \frac{243+8 {\tilde  z}^2 \left[  81 + 8 {\tilde  z}^2 \left(  171 + 78 {\tilde  z}^2 + 4 {\tilde z}^4 \right) \right]}{36 {\tilde  z}^2 \left( 3 + 4 {\tilde z}^2 \right)^2} \right\} v_2 \nonumber\\
& &  - \frac{16 {\tilde  z} \left( - 9 + 24 {\tilde z}^2 + 16 {\tilde z}^4 \right)}{3 \left( 3 + 4 {\tilde z}^2 \right)^2} v_1~,
\label{bulknum}
\end{eqnarray}
where prime ($^\prime$) denotes differentiation with respect to ${\tilde z}$, and where the subscript $1$ ($2$) corresponds to the field $\phi$ ($T$).  As we discussed in Section \ref{scalar}, to fully determine a mode we must specify the five quantities ($2N+1$, with $N=2$)
\begin{equation}
v_1 \vert_{\tilde \mu} \;,\;\; v_2 \vert_{\tilde \mu} \;,\;\;
v_1' \vert_{\tilde \mu} \;,\;\; v_2' \vert_{\tilde \mu} \;,\;\; {\widetilde m}_n^2~,
\label{5in}
\end{equation}
where we recall that ${\tilde \mu}$ is the position of the UV brane in the rescaled variable ${\tilde z}$.

We assume a stiff brane potential on the UV brane, which enforces the Dirichlet boundary conditions $\delta \phi = \delta T = 0$ (see Subsection \ref{stiff}). This in turn results in the two boundary  conditions
(\ref{v-bc-stiff}) and (\ref{bc-stiff}). Moreover, we fix the arbitrariness of the overall normalization by fixing the value of $v_2 =1$ at the UV brane. This, together with Eq. (\ref{v-bc-stiff}) gives
\begin{equation}
v_1 \vert_{\tilde \mu} = - \frac{2}{3} \, {\tilde \mu}\,, \;\;\;\; 
v_2 \vert_{\tilde \mu} = 1~.
\label{norma}
\end{equation}

We are left with the three parameters $v_1' \vert_{\tilde \mu} ,\; v_2' \vert_{\tilde \mu} ,\; {\widetilde m}_n^2$ subject to the constraint (\ref{bc-stiff}):
\begin{eqnarray}
&&   v_1' \vert_{\tilde \mu} - \frac{3}{2 \, {\tilde \mu}}  \,  v_2' \vert_{\tilde \mu}  = - \frac{{\tilde m}_n^2}{8 \, {\tilde \mu}^2} \left( 3 + 4 \, {\tilde \mu}^2 \right) \nonumber\\
 && \quad\quad\quad\quad + \frac{1}{12 \, {\tilde \mu}^2} \left( 9 + 16 \, {\tilde \mu}^4 \right)
+ 4 + \frac{8}{3 + 4 \, {\tilde \mu}^2} \,\,.
\label{getvp1}
\end{eqnarray}
We used this constraint to determine $ v_1' \vert_{\tilde \mu} $ as a function of $v_2' \vert_{\tilde \mu} $
and ${\widetilde m}_n^2$.

The bulk coordinate ${\tilde z}$ extends to $+\infty$. To understand the role of the associated boundary conditions, we studied the bulk equations in the limit of large ${\tilde z}$: 
\begin{eqnarray}
v_1'' & = & \left[ 4 \mu^4 z^2 + {\cal O } \left( z^0 \right) \right] v_1 + \left[ - \frac{16 \mu^3}{3} z + {\cal O } \left( z^{-3} \right) \right] v_2~, \nonumber\\
v_2'' & = &  \left[ - \frac{16 \mu^3}{3} z + {\cal O } \left( z^{-3} \right) \right] v_1 + \left[  \frac{4 \mu^4}{9} \, z^2 + {\cal O } \left( z^0 \right) \right] v_2~.\nonumber\\
\label{bulk-largez}
\end{eqnarray}
We obtained the approximate  solutions to these equations under the assumption that one mode is significantly larger than the other one in this limit; the subdominant mode can then be disregarded in the equation of the dominant mode; this equation can be solved analytically, and we can then insert this solution in the remaining equation to obtain the subdominant mode. We then studied the large ${\tilde z}$ limit of the solutions, and verified that the starting assumption (the subdominant mode can be neglected in the equation of motion of the dominant one) holds. In this way, we obtained four solutions, that form a complete basis for the solutions of the bulk equations. At ${\tilde z} \gg 1$, the four solutions read
\begin{eqnarray}
v_2 &\simeq& - \frac{3 C_1}{2  {\tilde  z}^{3/2} } \, {\rm e}^{-{\tilde z}^2} 
 \ll v_1  \simeq  \frac{ C_1 }{ \sqrt{\tilde z} }\, {\rm e}^{-{\tilde z}^2 }~, \nonumber\\
v_2 &\simeq& - \frac{3 D_1}{2 {\tilde z}^{3/2} } \, {\rm e}^{{\tilde z}^2} 
 \ll v_1  \simeq  \frac{ D_1 }{ \sqrt{\tilde z} }\, {\rm e}^{{\tilde z}^2}~, \nonumber\\
v_1 &\simeq&  \frac{3 C_2}{2 {\tilde z}^{3/2} } \, {\rm e}^{-\frac{1}{3} {\tilde z}^2} 
 \ll v_2  \simeq  \frac{ C_2 }{ \sqrt{\tilde z} }\, {\rm e}^{-\frac{1}{3} {\tilde z}^2}~,    \nonumber\\
v_1 &\simeq&  \frac{3 D_2}{2 {\tilde z}^{3/2} } \, {\rm e}^{\frac{1}{3} {\tilde z}^2} 
 \ll v_2  \simeq  \frac{ D_2 }{ \sqrt{\tilde z} }\, {\rm e}^{\frac{1}{3}  {\tilde z}^2}~. 
\label{modes-largez}
\end{eqnarray}
We note that two of these solutions are exponentially decreasing at large ${\tilde z}$, while the other two are exponentially increasing.

As we discussed in the previous Section, at this stage of the computation we can only guess some values for the initial parameters $v_2' \vert_{\tilde \mu} ,\; {\widetilde m}_n^2$. This guarantees that the boundary conditions at the UV brane are satisfied. If we start from any point in this two dimensional space of initially guessed parameters, and solve  the bulk equations (\ref{bulknum}) from the UV brane to the asymptotic  ${\tilde z} \gg 1$ region, we obtain solutions whose large ${\tilde z}$ asymptotics is  a linear combination of the four modes (\ref{modes-largez}). Only linear combinations that have $D_1 = D_2 = 0$ correspond to a normalizable, and hence physical, Kaluza-Klein mode. Given that we need to satisfy two conditions, and the space of initial parameters has dimension two, we expect that only a discrete set of points in this space corresponds to a physical mode.  Our goal is to identify these  points.

As can be expected, none of the initial guesses corresponds to a physical mode, and therefore we resorted to the following numerical algorithm. For the present discussion, let us denote by $\alpha \equiv {\widetilde m}_n^2 $ and $\beta \equiv v_2' \vert_{\tilde \mu} $ the two coordinates in the space of initial parameters, and by $f \left( \alpha ,\, \beta \right)$ and  $g \left( \alpha ,\, \beta \right)$ the values of the two wavefunctions $v_1$ and $v_2$, respectively, at some large value 
${\tilde z}_{\rm end}$. Physical modes correspond to zeros of these two functions. We numerically searched for zeros using Newton's method: starting from an initial guess $\left\{ \alpha_0 ,\, \beta_0 \right\}$ we initiate a succession $\left\{ \alpha_n ,\, \beta_m \right\}$ that, if it is convergent, converges to a zero of $f$ and $g$. The iteration step of the succession is
\begin{eqnarray}
\alpha_{n+1} & = & \alpha_n + \frac{f_{,\beta} \, g - g_{,\beta} \, f}{f_{,\alpha} \, g_{,\beta} - f_{,\beta} \, g_{,\alpha}}~, \nonumber\\
\beta_{n+1} & = & \beta_n + \frac{g_{,\alpha} \, f - f_{,\alpha} \, g}{f_{,\alpha} \, g_{,\beta} - f_{,\beta} \, g_{,\alpha}}~, 
\label{newton}
\end{eqnarray}
where the subscript comma denotes differentiation using finite differences, namely $f_{,\alpha} \equiv \frac{f \left( \alpha + \epsilon ,\, \beta \right) - f \left( \alpha - \epsilon ,\, \beta \right) }{2 \epsilon} $, and similarly for the other derivatives. The iteration step is obtained by Taylor expanding  
$f \left( \alpha_{n+1} ,\, \beta_{n+1} \right)$ and $g \left( \alpha_{n+1} ,\, \beta_{n+1} \right)$ in  terms of  $f ,\, g$ and their first derivatives, evaluated at $\left\{ \alpha_n ,\, \beta_n \right\}$. The expressions (\ref{newton}) are the algebraic solutions of these equations, after setting
 $f \left( \alpha_{n+1} ,\, \beta_{n+1} \right) =  g \left( \alpha_{n+1} ,\, \beta_{n+1} \right) = 0 $. 

Ideally, the algorithm converges to real solutions only for ${\tilde z}_{\rm end} = \infty$. In practice however, the numerical problem we are solving is  rather challenging. Indeed, we need to identify exponentially decreasing solutions among exponentially growing ones. No numerical solution starts from  values of  $\left\{ \alpha ,\, \beta \right\}$  that exactly correspond to a solution. The discrepancy, however small, necessarily results in an exponentially growing solution at ${\tilde z} \gg 1$.
The larger  the value of ${\tilde z}_{\rm end}$, the smaller the initial discrepancy needs to be, if one hopes that  the growing mode is still subdominant at   ${\tilde z}_{\rm end}$. After some trials, we found that values of 
${\tilde z}_{\rm end} \simeq 5$ result in successions that converge after $\sim 10^2$ or $ \sim 10^3$ iteration steps of (\ref{newton}). For larger values of  ${\tilde z}_{\rm end}$, the successions typically keep spanning large areas of the $\left\{ \alpha ,\, \beta \right\}$ plane without showing any sign of convergence.
We note that ${\tilde z}_{\rm end} \simeq 5$ are the values for which the exponential suppression/enhancement in (\ref{modes-largez}) starts to be significant.

\begin{figure}[h]
\centerline{
\includegraphics[width=0.35\textwidth,angle=-90]{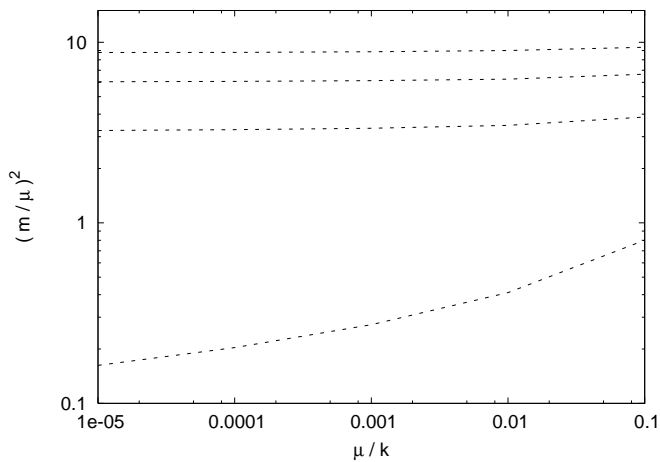}
}
\caption{Values of the lightest  eigenmasses as a function of ${\tilde \mu}=\mu/k$.
}
\label{fig:change-mu} 
\end{figure}

We generally expect that the effect of having a finite value for ${\tilde z}_{\rm end}$, rather than ${\tilde z}_{\rm end} = \infty$, is the following: any zero of $f$ and $g$ that we find arises because the exponentially  decreasing and the exponentially growing modes of (\ref{modes-largez}) provide an equal contribution to the wave function at ${\tilde z} \simeq {\tilde z}_{\rm end}$ (if this was not the case, the wave function would exponentially decrease, and not vanish, at ${\tilde z}_{\rm end}$). For 
sufficiently large ${\tilde z}_{\rm end}$, this however implies that the growing mode is extremely subdominant at the UV brane, and that the initial choice of  $v_2' \vert_{\tilde \mu} ,\; {\widetilde m}_n^2$ is very close to the physical one. For instance, in obtaining the eigenmasses that we show in Figure 
\ref{fig:change-mu}, we tried to push  ${\tilde z}_{\rm end}$ to the highest possible values that allowed for a convergence of (\ref{newton}), and we verified that a small decrease of  ${\tilde z}_{\rm end}$ from this highest possible value did not significantly change the values of  $v_2' \vert_{\tilde \mu} ,\; {\widetilde m}_n^2$ to which (\ref{newton}) converged to.

Our main goal in the current example  is to study whether  the soft wall model \cite{Batell:2008zm} is stable. Clearly, our numerical search will never produce a definite proof of this, since one may in principle think that modes with negative $m^2$ exist, but our numerical scheme could not find them. However, as we now discuss, our search algorithm converged to more than $100$ eigenmodes with ${\widetilde m}_n^2 > 0$, and there were no modes with  ${\widetilde m}_n^2 < 0$. We believe that this provides substantial evidence that the soft wall model is stable.

We performed two extensive searches, with different values of ${\tilde \mu}^2$. In the first search, we fixed ${\tilde \mu} = 0.01$, and  ${\tilde z}_{\rm end}  = 5$, and we took a grid of $250 \times 500$ values in the plane of initial values; specifically, the values of ${\widetilde m}_n^2$ ranged from $-49.9$ to $-0.1$ in steps of $0.2$, and  $ v_2' \vert_{\tilde \mu} $ from $-49.9$ to $49.9$ in steps of $0.2$ (starting from negative values of  ${\widetilde m}_n^2$ 
so as to maximize the chance of finding some tachyonic solution, if it existed). Starting from each of the $125,000$ points in the grid, we performed $400$ iterations of (\ref{newton}). The iterations converged to $118$ final points with ${\widetilde m}_n^2 > 0$, and with $\sqrt{f^2+g^2} \ll 1$ (therefore, they correspond to stable solutions of the problem). In the second search, we fixed  ${\tilde \mu} = 0.001$, and  ${\tilde z}_{\rm end}  = 5$.  We took a grid of $100 \times 200$ values in the plane of initial values; specifically, the values of ${\widetilde m}_n^2$ ranged from $-59.7$ to $-0.3$ in steps of $0.6$, and  $ v_2' \vert_{\tilde \mu} $ from $-597$ to $11343$ in steps of $60$. We performed $400$ iterations of (\ref{newton}), which  converged to $67$ final points,  with ${\widetilde m}_n^2 > 0$, and with $\sqrt{f^2+g^2} \ll 1$.

\begin{figure}[h]
\centerline{
\includegraphics[width=0.35\textwidth,angle=-90]{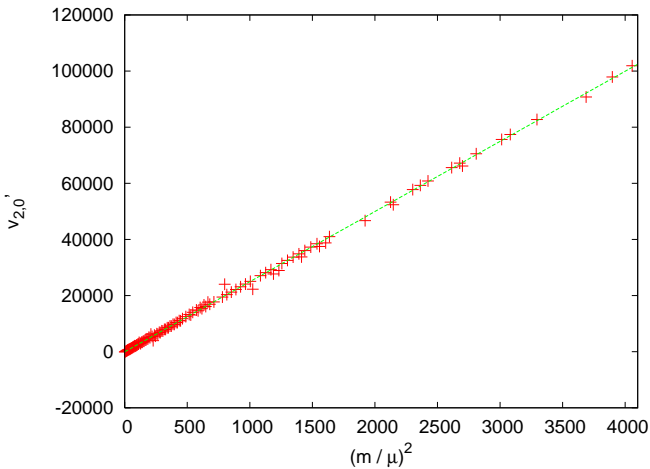}
}
\caption{Values of ${\widetilde m}_n^2 \equiv m_n^2 / \mu^2$ and $  v_{2,0}' \equiv  v_2' \vert_{\tilde \mu}  $ to which the algorithm described in the main text converged (with ${\tilde \mu} = 0.01$).  Note that all the solutions have positive mass-squared, ${\widetilde m}_n^2$. Also shown is the linear fit along which the parameters of the solutions are approximately distributed.}
\label{fig:converge} 
\end{figure}

The values of ${\widetilde m}_n^2$ and $ v_2' \vert_{\tilde \mu} $ of the convergence points are shown in Figure \ref{fig:converge}, for the search with ${\tilde \mu} = 0.01$. We see from the Figure  that many of the solutions have a final value of these parameters outside the range of the starting grid. This shows that the algorithm is able to find solutions in a  broad region of the  parameters.  We see that the  values  shown in the figure are roughly distributed along the line, $ v_2' \vert_{\tilde \mu} \simeq 25  {\widetilde m}_n^2 - 42$, although some scatter is present (for the search with  ${\tilde \mu} = 0.001$, we found that the parameters of the solution are roughly distributed along the line  $ v_2' \vert_{\tilde \mu} \simeq 250  {\widetilde m}_n^2 - 520$).  We also notice that the solutions we have found are not equally spaced along this line: this is a clear sign that the search has not found all the modes with the parameters in the range shown in the Figure. Particularly, it is natural to expect that the method failed to obtain modes with values of  ${\widetilde m}_n^2$ and $ v_2' \vert_{\tilde \mu} $ very different from those in the initial grid.  While the exact spectrum of the   soft wall model \cite{Batell:2008zm} is by itself a very interesting subject worth studying, this is not the goal of the present work. The main purpose of Figure \ref{fig:converge}  is to show that the algorithm that we have implemented did not find any unstable (${\widetilde m}_n^2 < 0$) mode. 

\begin{figure}[h]
\centerline{
\includegraphics[width=0.35\textwidth,angle=-90]{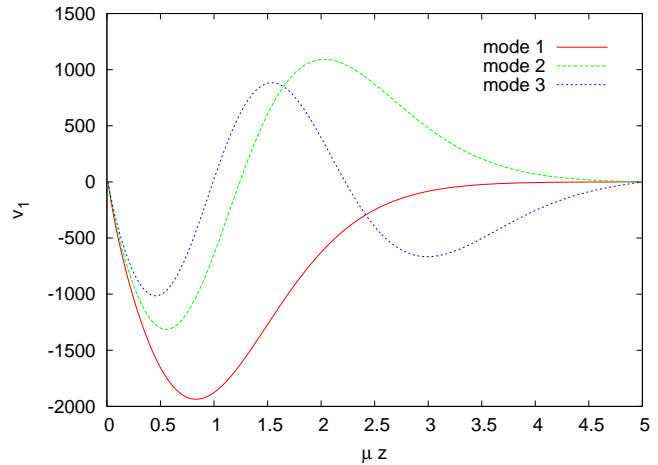}
}
\caption{Bulk wavefunctions of $v_1$ for the three lightest scalar Kaluza-Klein modes  for ${\tilde \mu} = 0.01$. The corresponding eigenmasses are ${\widetilde m}^2 \simeq 0.41 ,\, 3.47 ,\, 6.25$. 
 }
\label{fig:3modes-v1} 
\end{figure}

\begin{figure}[h]
\centerline{
\includegraphics[width=0.35\textwidth,angle=-90]{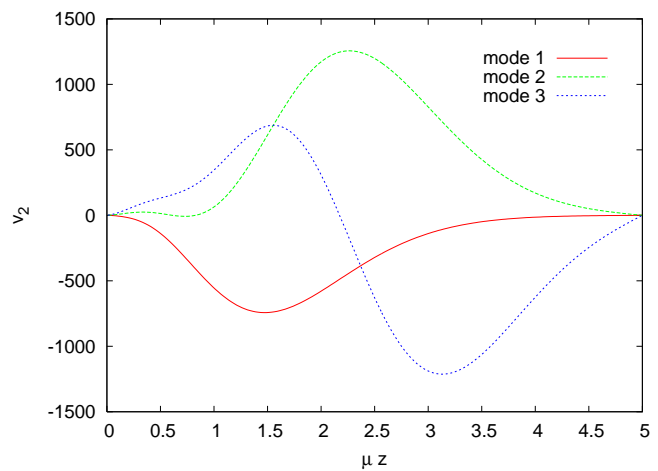}
}
\caption{Bulk wavefunctions of $v_2$ for the same modes appearing in 
Figure~\ref{fig:3modes-v1}. 
}
\label{fig:3modes-v2} 
\end{figure}

\begin{figure}[h]
\centerline{
\includegraphics[width=0.35\textwidth,angle=-90]{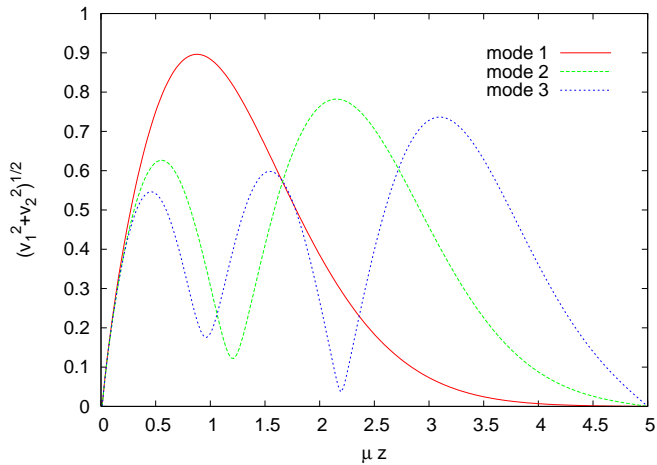}
}
\caption{Properly normalized bulk profile of the  same modes appearing in 
Figures~\ref{fig:3modes-v1} and \ref{fig:3modes-v2}. 
}
\label{fig:3modes-amplitude} 
\end{figure}

In Figures \ref{fig:3modes-v1} and  \ref{fig:3modes-v2}  we show the bulk profile of $v_1$ and $v_2$, respectively, of the three lightest modes obtained from this search. These wavefunctions are shown with the  ``provisory'' normalization  set by eq. (\ref{norma}). In Figure \ref{fig:3modes-amplitude} we show instead the bulk profile $\sqrt{v_1^2 + v_2^2}$ of the three modes after they have been properly normalized (see Subsection \ref{kin-action}). As typical, the profile of modes of increasing mass presents an increasing number of maxima and minima, and extends further in the bulk.

The hierarchy in the model (namely, the ratio between the IR and the UV mass scale) is given by ${\tilde \mu}$. While a solution to the hierarchy problem requires ${\tilde \mu} \simeq 10^{-16}$, the two searches we have performed are limited to very moderate hierarchies, namely ${\tilde \mu } = 0.01$ and $0.001$. However, we also studied how the lightest eigenmasses scale with ${\tilde \mu}$, and we believe that our results can be extrapolated to small hierarchies.  In Figure  \ref{fig:change-mu} we 
present the value of four light eigenmasses  for ${\tilde \mu}$ as small as $10^{-5}$. We find that the lightest eigenmode  shown behaves differently than the heavier ones. For all these other modes,  the ratio ${\widetilde m}_n^2 = m_n^2 / \mu^2$ approaches a constant value at small ${\tilde \mu}^2$. Instead, we find that  the numerical values obtained for the lightest eigenmass  are extremely well approximated by
\begin{equation}
\frac{m_{\rm radion}^2}{\mu^2} \simeq \frac{1.9}{{\rm log } \left( \frac{k}{\mu} \right)}~.
\label{radionmass}
\end{equation}
We were actually able to obtain this eigenmass for values of ${\tilde \mu}$ as small as $10^{-10}$ (for smaller values, the algorithm was no longer able to converge to the solution), and we verified that the fit (\ref{radionmass}) continues to be valid  also at  such low ${\tilde \mu}$.

In eq.~(\ref{radionmass}) we referred to this mode as a radion, because the behavior of its mass is consistent with that of the radion in the Goldberger-Wise 
\cite{Goldberger:1999uk} stabilization mechanism, while the behavior of the heavier masses is consistent with the behavior of the higher KK modes in  
\cite{Goldberger:1999uk}. We recall that $\mu$ is the IR scale of the model, while $k$ is associated with the UV scale. We therefore find that the wave function of the radion and of the higher KK modes are peaked at values of  the bulk coordinate $z$ of the order of the inverse IR scale. Moreover, we find that the higher KK modes obtain a mass of order the IR scale, while the radion mass is suppressed with respect to this value by the logarithm of the ratio between the UV and the IR scale. These  properties coincide with those obtained  in the  Goldberger-Wise \cite{Goldberger:1999uk} stabilization mechanism, as shown in the  analysis of \cite{Csaki:2000zn,Medina:2010mu} (in the computation of \cite{Csaki:2000zn}, the logarithmic suppression is encoded in the factor $u/k \sim 1/\left( k \, r_0 \right) \sim 1/  {\rm log } \left( {\rm UV} / {\rm IR} \right)$).

The nature of these modes as IR modes is also confirmed by the computation of their normalization, that we performed  for the specific case of  ${\tilde \mu} = 0.01$. Specifically, we evaluated $C_{nn}$ in eq. (\ref{Cmn}) for the four modes of   Figure \ref{fig:change-mu}. We found that the contribution to $C_{nn}$  from the bulk integral is about $5$ orders of magnitude greater than the contribution from the UV boundary at ${\tilde z} = {\tilde \mu}$. Since the wavefunctions $v_i$ (and, thus, their bulk integral) are peaked at values of $z$ of the order of the inverse IR scale, this confirms the IR nature of the modes. In an analytic computation, one would expect no contribution to $C_{nn}$  from the boundary at $z = \infty$. In our numerical computation, the contribution from the IR ``fictitious'' boundary at ${\tilde z}_{\rm end} = 5$ ranges from being $5$ to $7$ orders of magnitude smaller than the bulk integral. This is a measure of the goodness of our numerical results. 

A second check on our numerics is provided by the orthogonality of the modes. Ideally, the modes should be perfectly orthogonal to each other: $C_{mn} = 0$ for $m \neq n$. We find that $s_{mn} \equiv C_{mn} / \sqrt{C_{mm} \, C_{nn}}$ for $m \neq n$ is of  ${\rm O } \left( 10^{-5} \right)$ for the four  modes of  Figure \ref{fig:change-mu}. As a final check of the numerics, we verified that the wave functions $v_i$ of the four modes of Figure 
\ref{fig:change-mu}  is in good agreement with the analytic large ${\tilde z}$ solution that we wrote in the third line of eq. (\ref{modes-largez}). We recall that normalizable modes must behave as linear combinations of the first and third line of  (\ref{modes-largez}) at large ${\tilde z}$. Since the mode in the third line ($\propto {\rm e}^{-{\tilde z}^2/3}$) decreases less than the mode in the first line  ($\propto {\rm e}^{-{\tilde z}^2}$), only the ${\rm e}^{-{\tilde z}^2/3}$ component will be visible if the mode is  a  generic linear combination of these two solutions at large ${\tilde z}$.

\section{Conclusion} \label{conclusions}

The main goal, and result of this  work is to provide a formalism to study the perturbations, and the stability of codimension one brane compactifications with $N$ bulk scalar fields. We obtained the closed set of equations for such perturbations, and arranged them in an explicit eigenvalue problem. These equations are valid for arbitrary bulk and brane potentials for the scalar fields, and  therefore can  be readily employed to study any such configuration. Specifically, we identified the canonical perturbations of the system, eqs. (\ref{def-v}). They satisfy the second order bulk equations (\ref{eq-v}). Such equations must be supplemented by boundary conditions. In Section~\ref{generalU}, and in the following two Subsections, we provided such conditions for the case of generic brane potentials, infinitely stiff brane potentials, and no brane, respectively. As an example, we studied the system of perturbations in the soft-wall model of \cite{Batell:2008zm}, which is characterized by two bulk scalar fields. We found the scalar perturbations in this model  behave identically to those of the Goldberger-Wise stabilization mechanism \cite{Goldberger:1999uk}; namely, there is tower of KK modes with mass at the IR scale, and a lighter radion mode, whose mass is suppressed with respect to the KK tower by a large logarithm; all these modes are peaked at IR values of the bulk coordinate. Note that our formalism assumes that there are no massless scalar modes, but can be straightforwardly generalized to 
study examples in this class.

\vskip.25cm
\noindent{\bf Acknowledgements:} 
The work of TG is supported by the Australian Research Council. TG also 
thanks the SITP at Stanford for support and hospitality during the completion of this work.
The work of MP was supported in part by
DOE grant DE-FG02-94ER-40823 at the University of Minnesota.

\renewcommand{\theequation}{A-\arabic{equation}}
\setcounter{equation}{0}  

\section*{APPENDIX A: Brane boundary conditions}

We give the definitions and intermediate steps relevant for the computation of the brane junction conditions, mostly following the notation of \cite{Frolov:2003yi}. The brane position in the bulk is characterized by the vector $x^A \left( y^\mu \right)$, denoting the bulk coordinate  $x^A$ of a point identified by the position $y^\mu$ on the brane. This defines the basis vector  $e_{(\mu)}^A \equiv \frac{\partial x^A}{\partial x^\mu} $. The normal of the brane is then defined by the requirement that it is orthogonal to the basis vectors, and normalized to one:  $e_{(\mu)}^A \, n_A = 0 $ and $n_A n^A  = 1$.
The induced metric $\gamma_{\mu \nu}$ and the extrinsic curvature $K_{\mu \nu}$ are then
\begin{equation}
\gamma_{\mu\nu} = e_{(\mu)}^A \, e_{(\nu)}^B \left[ g_{AB} - n_A \, n_B \right]~,
\;\;\;\;
K_{\mu\nu} =  e_{(\mu)}^A \, e_{(\nu)}^B \, \nabla_A n_B~.
\end{equation}

We are interested in the results up to first order in the perturbations. We use the metric (\ref{metric-modes}), in the $E = B = E_\mu = 0$ gauge. We denote the brane position in the extra space as  $z_{\rm  brane } = z_{\rm background \; position} +  \zeta \left( x^\mu \right)$, where  $\zeta$ is a perturbation.
In vector notation, using the first component for the ordinary coordinates ($A=0,\dots,3$), and the final component for the extra coordinate ($A=5$), we have $e_{(\mu)}^A = \left( \delta_\mu^A ,\, \zeta_{,\mu} \right) $ and $ n_A = A \left( - \delta \zeta_{,\mu} \, \delta_A^\mu ,\, 1 + \Phi \right) $. This gives
\begin{equation}
\gamma_{\mu\nu} = A^2 \left[ \left( 1 + 2 \Psi \right) \eta_{\mu \nu} + h_{\mu \nu} \right]~,
\label{gamma-sol}
\end{equation}
and
\begin{eqnarray}
K_{\mu\nu} & = & \left[ A' \left( 1 - \Phi + 2 \Psi \right) + A \, \Psi' \right] \eta_{\mu\nu} - A \, \zeta_{,\mu \nu} \nonumber\\
& - & \frac{A}{2} \left( B_{\mu,\nu} + B_{\nu,\mu} \right) + \frac{A}{2} \, h_{\mu \nu}' + A' \, h_{\mu \nu}~.
\label{K-sol}
\end{eqnarray}

\section*{APPENDIX B: Scalar equations in the bulk}

We linearize the bulk equations (\ref{einstein-eq}) and (\ref{scalar-eq}) in the scalar perturbations (\ref{metric-modes}), in the $E=B=0$ gauge. As mentioned in the main text, the off diagonal $\mu\nu$
Einstein equations enforce $\Psi = - \Phi/2$, which we use to eliminate $\Psi$. Among the remaining equations, the $\mu 5$ Einstein equation is already given in eq.~(\ref{con2}) of the main text.
One can then show that the part of the $\mu\nu$ equation proportional to $\eta_{\mu \nu}$ (the one with the ellipsis in eq.~(\ref{con0})) can be written as a combination of  (\ref{con2}) and its $z$ derivative. We can therefore disregard it.

We are left with eq.~(\ref{scalar-eq}) and the $55$ component of  (\ref{einstein-eq}), which are, respectively,
\begin{eqnarray}
 && - \Box \delta \phi_i - \delta \phi_i'' - \frac{3 A'}{A} \delta \phi_i' \nonumber\\
&&\quad\quad\quad + 3 \phi_i' \Phi' + 2 A^2 V_{,i} \Phi +A^2 V_{,ij} \delta \phi_j = 0~, \nonumber\\
 && - \Box \Phi  +  \frac{2 \phi_k' \phi_k'}{3 M^3}  \Phi \nonumber\\
&&\quad\quad\quad - \frac{2}{3 M^3} \left(  \phi_k' \delta \phi_k' + \frac{4 A'}{A} \phi_k' \delta \phi_k - A^2 V_{,k} \delta \phi_k \right) = 0~, \nonumber\\
\label{bulk-sca-i55}
\end{eqnarray}
where we have used (\ref{con2}) to simplify the second of these equations. The left hand side of these equations (\ref{bulk-sca-i55}) will be denoted as ${\rm Eq}_i$ and $ {\rm Eq}_{55}$, respectively and also the left hand side of eqs. (\ref{eq-v}) are denoted as 
$\widetilde{\rm Eq}_i $. Using (\ref{con2}) repeatedly, one can show that
\begin{eqnarray}
\frac{1}{2} \frac{d}{d z} {\rm Eq}_{55}  + \frac{ A'}{A} {\rm Eq}_{55} - \frac{1}{3 M^3} \phi_k' {\rm Eq}_k &=& 0~,
\nonumber\\
-\sqrt{2} A^{3/2} \left(  {\rm Eq}_i + \frac{A \, \phi_i'}{2 A'} \,  {\rm Eq}_{55} \right) + \widetilde{\rm Eq}_i &=& 0~.
\end{eqnarray}

The first relation indicates that one equation in the  $ \left\{ {\rm Eq}_{i} = 0 ,\, {\rm Eq}_{55} = 0 \right\}  $ 
system is redundant. The first and second relations together show that the   $ \left\{ {\rm Eq}_{i} = 0  ,\, {\rm Eq}_{55} = 0  \right\}  $ system  of equations  for the $N+1$  variables $ \left\{ \delta \phi_i ,\, \Phi \right\} $ is equivalent to the system of equations $ \left\{  \widetilde{\rm Eq}_i = 0 \right\} $ for the $N$ variables $ \left\{ v_i \right\} $. This is possible because   the variables  $ \left\{ \delta \phi_i ,\, \Phi \right\} $ are subject to the constraint  (\ref{con2}).

\section*{APPENDIX C: Scalar boundary conditions}

In this Appendix we summarize the algebra needed to rewrite the boundary condition (\ref{jun2}) into the expression (\ref{junction}). We recall that our goal is to write the boundary conditions as expressions of the form $v_i' = \sum_j c_{ij} v_j$.

We start by taking the $z$ derivative of the definition of $v_i$, eq. (\ref{def-v}). We then eliminate $\delta \phi'$ and $\Phi'$ from the resulting expression using (\ref{con2}) and (\ref{jun2}). Finally, we eliminate $\delta \phi_i$ using  (\ref{def-v}). This gives rise to the expressions,
\begin{eqnarray}
\label{vip}
&&
v_i' - \frac{3 A'}{2 A} v_i - \left( \frac{A \phi_i' \, \phi_j'}{3 M^3 A'} \pm \frac{A}{2} U_{,ij} \right) v_j 
\nonumber\\
&&
=\left( 2 \sqrt{2} A^{3/2} \phi_i' - \frac{A^{9/2} V_{,i}}{\sqrt{2} A'} \pm \frac{A^{7/2}}{2 \sqrt{2} A'} U_{,ij} \, \phi_j' \right) \Phi~,\nonumber \\
\end{eqnarray}
which are equivalent to the boundary conditions  (\ref{con2}), once we assume  that the modes satisfy the bulk equations in the bulk (which clearly must be the case). We need to eliminate $\Phi$ in favor of a linear combination of $v_i$. From the second equation of (\ref{bulk-sca-i55}), together with (\ref{con2}) and (\ref{def-v}), we find
\begin{eqnarray}
\Box \Phi &=& \frac{\sqrt{2}}{3 M^3 A^{3/2}} \Bigg[ \phi_k' v_k' - A^2 V_{,k} v_k 
\nonumber\\
&& +  \left( \frac{5}{2} \frac{A'}{ A} -\frac{A}{A'} \frac{\phi_k' \phi_k'}{3 M^3} \right) \phi_q' v_q \Bigg]~.
\label{Phi-v-vp}
\end{eqnarray}
In principle, we can solve and obtain $\Phi$ from this relation, and insert it in (\ref{vip}). The resulting equations would be functions of $v_i$ and $v_i'$ only, but would not be of the required form (the linear combination $\propto \phi_k' \, v_k'$ on the right hand 
side of (\ref{Phi-v-vp}), means that a nontrivial inversion is needed to solve for $v_i'$).

Instead to achieve our goal, $\Phi$ should be expressed as a linear combination of $v_i$ only. This can be obtained if there is a second relation between $\Phi$, $\phi_k' \, v_k'$, and $v_i$. This is readily achieved if we  multiply  (\ref{vip}) by $\phi_i'$, and sum over $i$:
\begin{eqnarray}
\phi_i' v_i' &=& \left( \frac{3}{2} \frac{A'}{A} + \frac{A \phi_k' \phi_k'}{3 M^3 A'} \right) \phi_i' v_i \pm \frac{A}{2} U_{,ij} \phi_i' v_j \nonumber\\ 
&&\!\!\!\!\!\!\!\!\!\!\!\!\!\!\! + \left( 2 \sqrt{2} A^{3/2} \phi_k' \phi_k' - \frac{A^{9/2} \phi_k' V_{,k}}{\sqrt{2} A'} \pm \frac{A^{7/2}}{2 \sqrt{2} A'} U_{,ij} \, \phi_i' \phi_j' \right) \Phi~.\nonumber\\
\end{eqnarray}
Substituting this expression into (\ref{Phi-v-vp}) gives rise to an equation containing only the modes $\Phi$ and $v_i$. This equation is then used to express $\Phi$ as a linear combination of $v_i$. Inserting this expression into (\ref{vip}) provides the boundary conditions (\ref{junction}) in the required form.

\end{document}